\documentstyle[epsf,epsfig]{elsart}
\input{psfig}
\newcommand{\pnk}{p~\rightarrow~\nu~K^+}
\newcommand{\km}{K^+~\rightarrow~\mu^+\nu_\mu}

\newcommand{\kpz}{K^+~\rightarrow~\pi^+\pi^0}
\newcommand{\etal}{{et~al.}}
\begin{document}
\begin{frontmatter}
\hoffset -.68in
\voffset -1.0in
\textwidth 6.5in
\textheight 9.0in
\title{\bf SEARCH FOR THE PROTON DECAY MODE $\pnk$ IN SOUDAN~2} 
\author[3]{W.W.M. Allison},
\author[4]{G. J. Alner},
\author[1]{D. S. Ayres},
\author[3]{G. Barr\thanksref{a}},
\author[6]{W. L. Barrett},
\author[2]{C. Bode},
\author[2]{P. M. Border},
\author[3]{C. B. Brooks},
\author[3]{J. H. Cobb},
\author[4]{R. Cotton},
\author[2]{H. Courant}, 
\author[2]{D. M. Demuth}, 
\author[1]{T. H. Fields},
\author[3]{H. R. Gallagher},
\author[4]{C. Garcia-Garcia\thanksref{b}}, 
\author[1]{M. C. Goodman}, 
\author[2]{R. Gran}, 
\author[1]{T. Joffe--Minor},
\author[5]{T. Kafka}, 
\author[2]{S. M. S. Kasahara},
\author[1]{W. Leeson}, 
\author[4]{P. J. Litchfield}, 
\author[2]{N. P. Longley\thanksref{c}},
\author[5]{W. A. Mann}, 
\author[2]{M. L. Marshak}, 
\author[5]{R. H. Milburn}, 
\author[2]{W. H. Miller}, 
\author[2]{L. Mualem},
\author[5]{A. Napier}, 
\author[5]{W. P. Oliver}, 
\author[4]{G. F. Pearce},  
\author[2]{E. A. Peterson},
\author[3]{D. A. Petyt\thanksref{d}}
\author[1]{L. E. Price}, 
\author[2]{K. Ruddick}, 
\author[5]{J. Schneps},  
\author[2]{M. H. Schub},  
\author[1]{R. Seidlein}, 
\author[3]{A. Stassinakis}, 
\author[5]{H. Tom}, 
\author[1]{J. L. Thron}, 
\author[2]{V. Vassiliev}, 
\author[2]{G. Villaume}, 
\author[2]{S. Wakeley}, 
\author[5]{D. Wall}, 
\author[3]{N. West}, 
\author[3]{U. M. Wielgosz}
%
%
%
\maketitle
%
\address[1]{Argonne National Laboratory, Argonne, IL
60439, USA }
\address[2]  {University of Minnesota, Minneapolis, MN
55455, USA }
\address[3]{ Department of Physics, University of Oxford,
Oxford OX1 3RH, UK }
\address[4]  {Rutherford Appleton Laboratory, Chilton, Didcot,
Oxfordshire OX11 0QX, UK }
\address[5]{ Tufts University, Medford, MA 02155, USA }
\address[6]  { Western Washington University, Bellingham, WA 98225, USA }

\thanks[a]  {Now at CERN, Geneva, Switzerland}
\thanks[b]  {Now at IFIC, E-46100 Burjassot, Valencia, Spain  }
\thanks[c]  {Now at Swarthmore College, Swarthmore, PA
19081, USA}
\thanks[d] {Now at Rutherford Appleton Laboratory, Chilton, Didcot,
Oxfordshire OX11 0QX, UK }

\begin{abstract}We have searched for the proton decay mode 
$\pnk$ using the one-kiloton Soudan~2 high resolution calorimeter.
Contained events obtained from a 3.56 kiloton-year fiducial 
exposure through June 1997 are examined for occurrence of a visible 
$K^+$ track which decays at rest into $\mu^+\nu$ or $\pi^+\pi^0$.
We found one candidate event consistent with 
background, yielding
a limit, 
$\tau/B({\pnk}) > 4.3 \times 10^{31}$ years 
at 90\% CL with no background subtraction.
\end{abstract}
\end{frontmatter}

\section{Introduction}
The prediction of many Grand Unified Theories, that 
nucleon decay occurs at accessible lifetimes, remains unverified 
but continues to motivate experimental searches.  
This expectation was
foreshadowed in part by Sakharov's early suggestion that 
the simultaneous effects of 
baryon number violation, C and CP violation, and departure from 
thermal equilibrium could produce the baryon-antibaryon  asymmetry 
observed in the universe.\cite{bib:sak1}
It is interesting and suggestive that no fundamental 
symmetry is known which implies the conservation of baryon number. 
Currently, nucleon decay as a consequence of the minimal SU(5) GUT 
model is considered to be ruled out 
experimentally\cite{bib:Hirata,bib:becker}.
However, other unification models, both with and without 
supersymmetry, predict baryon number violating processes.
Amplitudes for these processes involve the exchange of 
new particles with unknown 
masses so that precise nucleon lifetimes are not 
predicted.  The expectation that these masses will be in the range 
between the GUT scale of $\sim10^{15}$~GeV and the Planck mass of 
$\sim10^{19}$~GeV leads to proton lifetimes in the range 
$10^{32}-10^{35}$ years\cite{bib:hisano}.
Decay modes with strange particles such 
as $\pnk$, are 
usually favored in models which incorporate 
supersymmetry\cite{bib:wess,bib:nath}.
\par Previous searches for $\pnk$ have been reported by the IMB, 
Kamiokande and Frejus 
collaborations\cite{bib:Haines,bib:Hirata,bib:Berger}.
The $K^+$ track can be imaged in ionization 
calorimeters such as Soudan~2 and Frejus, 
but is usually below Cherenkov threshold in water.  
IMB searched for an excess of events in 
a region of anisotropy and energy with a large background\cite{bib:Haines}.  
Kamiokande looked for an excess of single ring 
mu-like events 
between 215 and 255 MeV/c with a muon decay, and also for 
three-ring events compatible with an invisible, stopped 
$\kpz$ decay\cite{bib:Hirata}.  
Frejus
used two-track events with ranges consistent 
with the $K^+$ and the $\mu^+$\cite{bib:Berger}.  
\par In the Soudan~2 analysis, we use 
both the visibility of the $K^+$ in a fine grained tracking 
calorimeter and the visibility of the decay electron from a 
stopped $\mu^+$ to reduce backgrounds from atmospheric neutrino interactions.
We searched for the proton decay mode $\pnk$ using two 
$K^+$ decay channels, $\km$ and $\kpz$.

\section{The Soudan~2 Detector}
\par  The Soudan~2 detector is a time projection, modular iron tracking 
calorimeter with a total mass of 974 metric tons and fiducial mass 
of 770 tons.  Details of module construction and performance 
may be found in References \cite{modcon,modperf,gallagher}.
The detector is assembled as a close-packed rectangular stack of 224  
modules;  each module is made from 1.6~mm thick sheets of corrugated
steel, stacked in a hexagonal ``honeycomb" structure.  The average 
density is 1.58~g/cm$^3$.
On the walls of the underground cavern surrounding the detector, there 
is an 
active ``veto" shield comprised of double-layer, hexagonal cell,
aluminum proportional tubes\cite{shield}.

\par Two million Hytrel plastic
drift tubes (1.0 m long by 15 mm in diameter) fill the spaces
in the honeycomb stacks.
Ionization electrons deposited in an Ar/CO$_2$ gas mixture
drift toward either end of 
the tube in a 180 volt/cm electric field with a velocity 0.8 
cm/$\mu$sec.
Upon reaching the tube end, the 
electrons are detected by vertical anode wires and horizontal 
cathode strips.  
Each crossing of a tube by an ionizing particle 
can create an anode/cathode signal at a common 
drift time which we call a ``hit".  The pulse area, which is 
proportional to the integrated charge deposited in the tube,
and the drift 
time for each hit 
are recorded by both the anode and cathode electronics.
The primary trigger requires at least 7 hits, separated by at 
least 600~ns, in a group of 16 anode channels or at least 8 hits 
in a group of 16 cathode channels within a $50$~$\mu$sec window.
The trigger efficiency for proton decay 
final states considered here is $\sim$ 85\%.
The complete detector triggers at a  rate of $\approx$ 0.5 
Hz from 
naturally occurring radioactivity and cosmic ray muons.
Every 240 seconds a ``pulser" trigger provides a snapshot 
of the random background levels in the main detector.  These are 
used as underlying events to add detector noise to 
Monte Carlo events.  

\section{Event Analysis}
\subsection{Overview}
\par The data analysis proceeds in three stages.  
First we identify ``contained events".  Event prongs are defined by scanning
as track-like ($\pi$, $\mu$ or p) or shower-like (e or $\gamma$).  Contained
events are defined as having no hits on tracks or the main body of showers
which are less than 20cm from the outside surface of the detector and the 
prongs do not start or end between modules.  This is the same contained
event selection as was used for our atmospheric neutrino analysis 
\cite{bib:allison}.
Studies in reference \cite{bib:allison} showed that the efficiency for correct 
identification  was 98\% for tracks and 94\% for showers.  An absence of shield
activity was required.
Second, the events are required to have a topology consistent with
the proton decay channel under study, based on counting the number 
of visible tracks and showers.
Finally, kinematic selections which characterize a particular 
proton decay mode are applied to the data and also to event 
samples which monitor background processes.
\par The analysis procedure involves finding efficient selection criteria 
using our proton decay Monte Carlo program, while minimizing the backgrounds 
from atmospheric neutrinos and atmospheric muons.  The former 
backgrounds are 
calculated using the atmospheric neutrino Monte Carlo program described in 
Reference \cite{bib:allison}, which incorporates the flux 
predictions of Barr, Gaisser 
and Stanev\cite{barr}.
Backgrounds from atmospheric muons 
may result when muons inelastically scatter in the rock outside the 
active shield.  
We use the term ``rock" event to describe the interactions of
a resulting secondary such as 
a neutron or $K_L$ which 
goes into the Soudan~2 calorimeter and causes
a contained vertex event.  Most 
rock events have hits in the Soudan~2 shield which are time-coincident 
with the 
contained event in the calorimeter.  These shield-tagged events 
are used to estimate any background from rock events without 
shield tags.  A detailed analysis of the penetration depth distributions 
of events with and without shield hits has led to the conclusion 
that $91.3\% \pm 3.7\%$ of all rock events have shield tags.
\label{sec:pdepth}
\par Our Monte Carlo simulation program tracks
the decay products through the detector geometry and 
generates electronic  hits in the same format as real data.  The generator 
starts with a parent (or target) nucleon within a nucleus.  The 
nucleon is considered 
to have a Fermi momentum chosen from the Bodek and Ritchie 
parameterization\cite{bodek}.  The spatial location and the atomic 
number of the parent 
nucleus is chosen according to the composition and mass 
distribution of the detector.   The rescattering of pions 
within a parent nucleus is generated according to a 
phenomenological model\cite{intranuke}.  Parameters of the model 
have been set to reproduce pion production by 
low energy neutrinos in $\nu_\mu$-Ne and $\nu_\mu$-deuteron 
reactions\cite{merenyi}.  
Inelastic intranuclear rescattering for $K^+$ mesons
is not expected to be significant due to the absence of low-lying 
$K^+N$ resonances and is not simulated.
The detector response used in this 
Monte Carlo program was verified against calibration data from the ISIS 
facility 
using $\pi$, e, $\mu$ and p beams at a variety of angles 
and energies\cite{garcia}.  
\subsection{Search for the Decay $\pnk, \km$}
\par
\begin{figure}[ht]
\vspace*{0.5in}
\epsfig{file=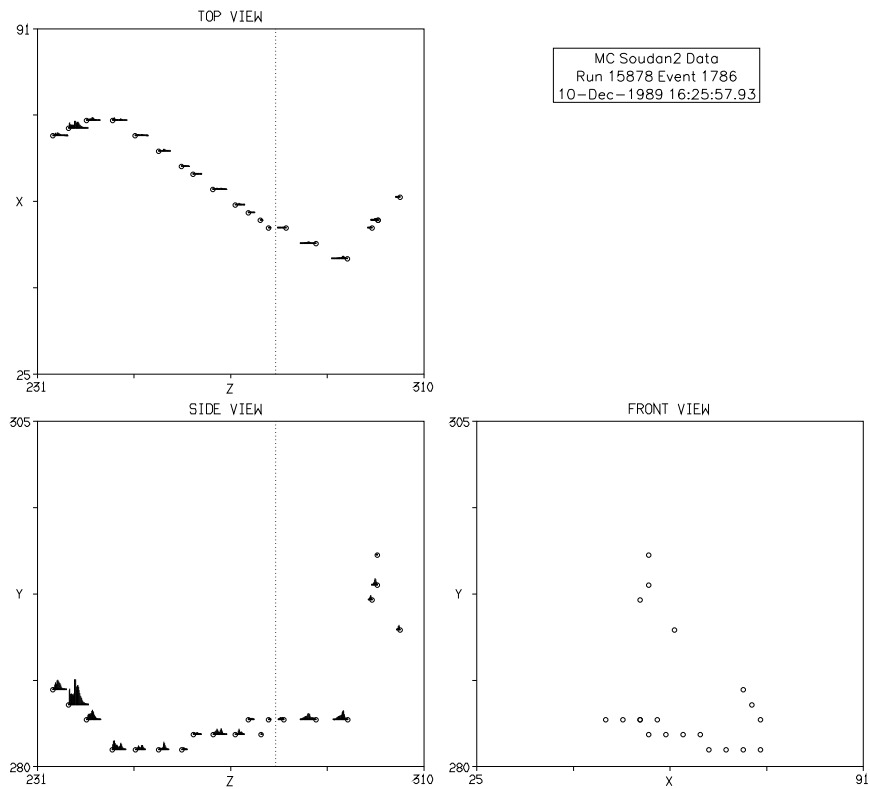,width=6in,angle=90}
\caption{\label{fig:pnk}
Monte Carlo event of $\pnk$. The three projected views are shown. 
The xz/yz views correspond to the anode/time and cathode/time 
views.  The xy view is based on matching anode and cathode hits
using their time and pulse shape.  For each hit, the pulse 
area is proportional to the recorded energy loss.  
The $K^+$ is the short
heavily ionizing track on the left/left/right of the xz/yz/xy 
plot.  Three hits from the $e^+$ decay of the $\mu^+$ appear
at the right/right/top.  Scales are in cm.}
\end{figure}

 To choose the selection criteria and determine efficiencies
we generated a large sample of $\km$ Monte Carlo events from 
proton decay.
Features of the typical Monte Carlo event shown in
Figure~\ref{fig:pnk} include a highly ionizing
$K^+$ which (usually) comes to rest, and emits a 236 MeV/c $\mu^+$.
The $\mu^+$ has an average range of 42 cm in the Soudan~2 
detector.  At the end of its range, the $dE/dx$ of the muon is 
rising.  The $\mu^+$ comes to rest and decays into an 
$e^+$, which gives an average
of 3 hits.   

The results of the analysis for the proton decay simulation, 
the neutrino simulation, the data, and the shield-tagged rock background are
given in Table~\ref{tab:effpnk}.  
The simulated events are generated 
in the entire mass of the Soudan~2 detector.  Efficiencies 
within the fiducial mass are calculated by dividing the fraction of 
events which pass a set of cuts by the ratio of the fiducial mass
to the total mass (0.79). 

\par 
The Monte Carlo events were first subjected to a simulated trigger, 
which 81 events
failed.  Both Monte Carlo and data events were processed through a filter
program which applied containment criteria, reducing cosmic ray muons by a
factor of more than $10^3$.  114 MC events, essentially all with hits 
outside the containment volume, were rejected.   The remaining events were 
then scanned by physicists to remove the remaining non-contained events,
mostly events starting or ending on module boundaries.
At this stage, 204 $\pnk, \km$ events remained,
compared with 367 data events without shield hits, 1008 
shield-tagged rock 
events, and 1923 events from the atmospheric neutrino Monte 
Carlo.  The data corresponds to 3.56 fiducial kiloton-years of 
exposure, compared to 20.24 kiloton-years for the atmospheric neutrino MC.

\par
\begin{figure}[ht]
\vspace*{0.5in}
\epsfxsize=6.0in\epsffile[0 97 529 673 ]{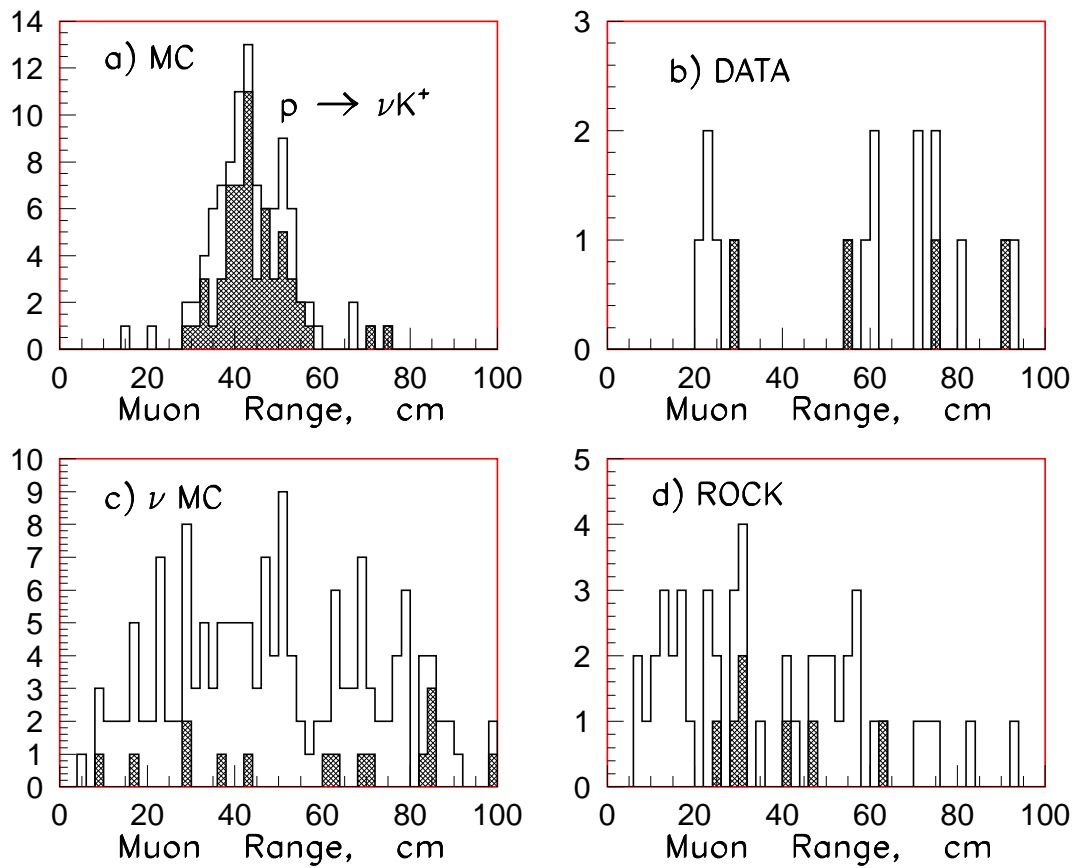}
\caption{\label{fig:muran}
Muon range distributions (before the muon decay cuts are imposed)
for a) 236 MeV/c $\mu$'s 
from the $\pnk$ Monte Carlo simulation, b) the data, c) the 
atmospheric neutrino Monte Carlo and d) the shield-tagged rock 
background.  The numbers of events (including overflows) are the same as in the 
row labeled ``$K$ range requirement" of Table 1.  The shaded events 
pass the muon decay cut.}
\end{figure}

 The required topology is two charged tracks with a common 
vertex.  Events in which both tracks appear to be protons based on
ionization and straightness are not included.
These topology features were exhibited by 95 of the remaining 204
events from the MC.  We also require that the $K^+$ candidate,  
which is usually the shorter
track, has a length less than 50 cm.   The muon track length 
distributions for all four event samples at this stage of the 
analysis are shown in 
Figure~\ref{fig:muran}.  
The range of muons from the $K$ decay is peaked at 43~cm whereas 
the background distributions are relatively flat.  The muon range 
was therefore required to lie
between 29 and 
58 cm.  Our final cut requires a visible muon decay electron having
two or more hits.  
This requirement discriminates strongly against neutrino induced 
background, since the predominant background is $\nu n \rightarrow 
\mu^- p$ and in our iron detector most $\mu^-$ are absorbed rather 
than decay after stopping.
\par After all cuts, our efficiency for accepting $\pnk, \km$ 
events in the 
fiducial volume is 14\%.  Two atmospheric $\nu$ Monte 
Carlo events pass all of the cuts and represent an expected 
background of 0.21 events, taking into account that we found 
the atmospheric neutrino flavor ratio ($\nu_\mu/\nu_e$)
to be only 0.61 of the expected 
value\cite{capri}.  The rock event background in the zero 
shield hit sample is calculated to be 0.19 events.  This was found 
based on the penetration depth analysis 
mentioned above, which found background in the track data equal to
9.5\% of 
the shield-tagged track sample.
\par As shown in Table~\ref{tab:effpnk}, one event in the data 
survives our cuts.

\begin{table}[ht]
\begin{center}
\begin{tabular}{|l|r|r|r|r|}                       \hline
Event Selection & PDK MC & $\nu$ MC & Rock & Data\\ \hline
MC decays in total detector & 493 & & & \\ \hline
Triggered detector & 412 & & & \\ \hline 
Containment filter & 298 & & & \\ \hline
Scanned as Contained & 204 & 1923 & 1008 & 367 \\ \hline 
Topology & 95 & 345 & 61 & 30 \\ \hline
$K$ range requirement & 95 & 286 & 54 & 27 \\ \hline
Muon range requirement & 88 & 62 & 21 & 1 \\ \hline
Visible muon decay & 55 & 2 & 2 & 1 \\ \hline
Exposure corrected background & & 0.21 & 0.19 & \\ \hline
\end{tabular}
\caption{Numbers of MC and data candidate 
events for 
$\pnk, \km$ which survive the triggering, containment, topology, and 
kinematic cuts of this analysis.  Events are generated in the full 
detector, while efficiencies in Table 3 are quoted for the 
fiducial volume.}
\label{tab:effpnk}.  
\end{center}
\end{table} 
\subsection{$\kpz$ analysis}
\par We searched for the mode $\pnk, \kpz$ by 
selecting events with a short heavily ionizing track
($K^+$ candidate), one other track and two showers.  
These features are illustrated in the Monte Carlo event shown in Figure
3.  The short highly ionizing track is the $K^+$ before it decays.
A $\pi^0$ is reconstructed from the two showers and the $K^+$ from the
$\pi^0$ and the second track, assumed to be a $\pi^+$.  At these low energies 
the two $\gamma$'s from the $\pi^0$ are usually identifiable.  The $K^+$
mass is required to be
in the range $100~$MeV/c$^2 < m_{K^+} < 
660~$MeV/c$^2$.  The $\pi^+$ and $\pi^0$ momenta are required to 
be in the ranges $80~$MeV/c$ < p_{\pi^+} < 400~$MeV/c and 
$40~$MeV/c$ 
< p_{\pi^0} < 390~$MeV/c.  The invariant mass of the two shower 
system is required to be in the range 
$10~$MeV/c$^2 < m_{2\gamma} < 290~$MeV/c$^2$.  
The resulting detection 
efficiency for this mode is 26\%.
\begin{figure}[ht]
\vspace*{0.5in}
\epsfig{file=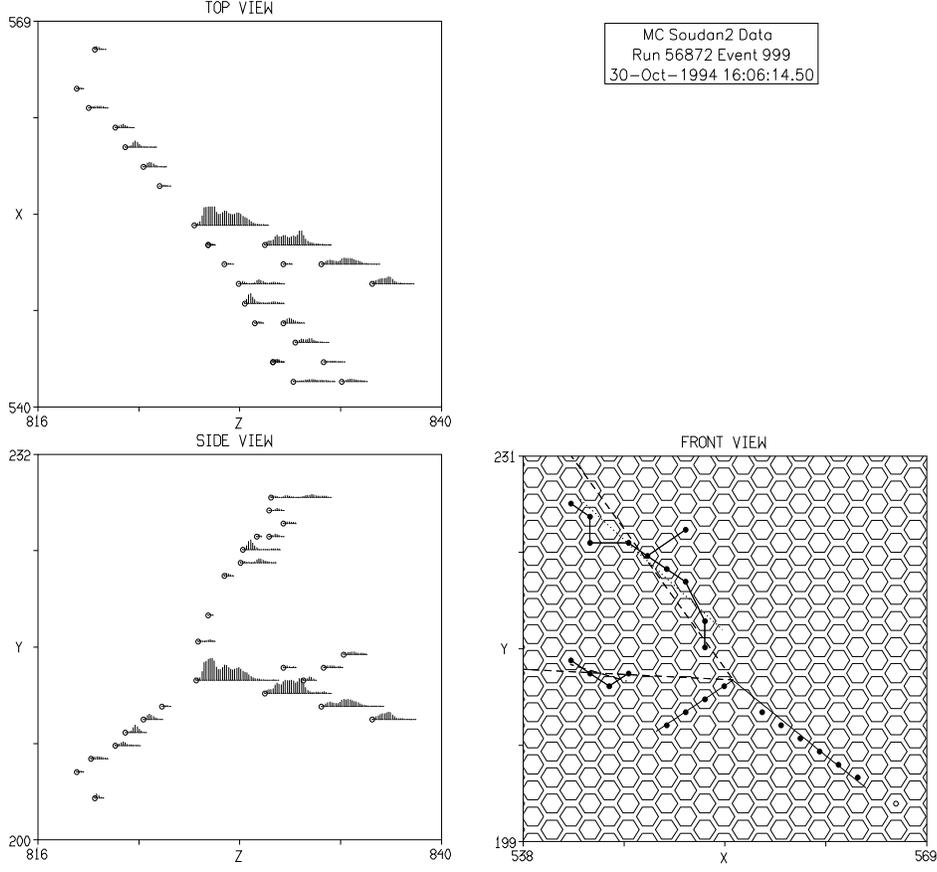,height=5in}
\caption{\label{fig:kpz}
Monte Carlo event of $\pnk;~\kpz$. 
The $K^+$ track of
four heavily ionizing hits, ranges and decays at rest, 
yielding
a pion track and two showers from the $K^+$ endpoint.  The 
showers
are the $\pi^0$ remnants; they appear in directions 
opposite to the
pion track and are overlapping in the xz view. The xy view also
shows the results of our track and shower fits.  Scales are in 
cm.}
                                                                
\end{figure}
\par The effects of these cuts on the proton decay simulation are 
shown in Table 2.  No data events pass these cuts.
 The 
background from atmospheric neutrino interactions is estimated to 
be 1.05 
events based upon our MC simulated $\nu$ events.  The background 
is from $\nu_e$ charged current and $\nu_\mu$ neutral current 
interactions, so the suppression factor of 0.62 is not used here.
The estimated 
background contribution  from rock events is 0.09 events, based 
on the one rock event which passed our cuts.  
\begin{table}[ht]
\begin{center}
\begin{tabular}{|l|r|r|r|r|}                       \hline
Event selection & PDK MC & $\nu$ MC & Rock & Data\\ \hline
MC decays in total detector & 493 & & & \\ \hline
Triggered detector & 442 & & & \\ \hline 
Containment filter & 317 & & & \\ \hline
Scanned as Contained & 229 & & & \\ \hline 
Topology & 106 & 18 & 3 & 5 \\ \hline
$K$ range requirement & 106 & 15 & 3 & 4\\ \hline
$K$ mass requirement & 106 & 11 & 3  & 1  \\ \hline
$\pi^+$  momentum cut & 103 & 7 &  2 & 0 \\ \hline
$\pi^0$ momentum cut & 103 & 6 & 1 & 0 \\ \hline
$\pi^0$ mass cut & 101 & 6 & 1 & 0 \\ \hline
Exposure corrected background & & 1.05 & 0.09 & \\ \hline
\end{tabular}
\caption{Numbers of MC and data candidate events for 
$\pnk, \kpz$ which survive the triggering, containment, topology, and 
kinematic cuts of this analysis.}
\label{tab:effkpz}.  
\end{center}
\end{table} 
\subsection{Limit calculation}
Using the two $K$ branching modes studied here we set a proton 
lifetime limit using the formula:
\begin{equation}
\frac{\tau}{B}( p \rightarrow \nu K^+) >  \frac{N_p \times T \times
[\varepsilon_1\times B_1(K) + \varepsilon_2\times B_2(K)]
}{\mu_1+\mu_2}  
\end{equation}
Here $N_p$ = 2.87 $\times 10^{32}$ is the number of protons in
a kiloton of the Soudan~2 detector, $T$~=~3.56~kiloton-years is the 
fiducial exposure $\varepsilon_i \times B_i(K)$ are the detection efficiencies 
quoted above times 
the appropriate $K^+$ decay branching fractions in 
Table~\ref{tab:effs}.  The $\mu_i$ are the constrained 90\% CL 
upper limits on the numbers of observed events, and are found by 
solving the equation 
\begin{equation}
0.10 = \frac{\sum_{n_1=0}^{n_{ev;1}}\sum_{n_2=0}^{n_{ev;2}} 
P(n_1,b_1+\mu_1) P(n_2,b_2+\mu_2)}
{\sum_{n_1=0}^{n_{ev;1}}\sum_{n_2=0}^{n_{ev;2}}
P(n_1,b_1) P(n_2,b_2)}
\end{equation}
with the constraint
\begin{equation}
\frac{\varepsilon_1 \times B_1(K)}{\mu_1} = \frac{\varepsilon_2 
\times B_2(K)}{\mu_2} =
\frac{\sum_{i=1}^{2} \varepsilon_i \times B_i(K)}{\sum_{i=1}^{2} 
\mu_i} 
\end{equation}
where $P(n,\mu)$ is the Poisson function, $e^{-\mu} \mu^n/n!$, and the 
$b_i$ are the estimated backgrounds.
With the one candidate observed for $\km$ and none found for 
$\kpz$, the values 
of $\mu$ in Table~\ref{tab:effs} are obtained, and the
combined limit without background subtraction 
is $4.3 \times 10^{31}$ years at 90\% CL.  With background 
subtraction, a limit $4.6 \times 10^{31}$ years is obtained.
\begin{table}[ht]
\begin{center}
\begin{tabular}{|l|l|l|}            \hline
 & $\km$ & $\kpz$ \\ \hline
$\varepsilon$ & 0.14 & 0.26 \\ \hline
$B(K)$ & 0.64 & 0.21 \\ \hline
$\varepsilon \times B(K)$ & 0.090 & 0.055 \\ \hline
$\mu(b=0)$ & 2.1 & 1.3 \\ \hline
b (background) & 0.4 & 1.1 \\ \hline
$\mu(b)$   & 2.0 & 1.2 \\ \hline
\end{tabular}
\caption{Summary of detection efficiencies, branching fractions, 
backgrounds, and combined upper limits at 90\% CL with and without 
background subtraction for the 
$\pnk$ channels analyzed.}
\label{tab:effs}.  
\end{center}
\end{table} 
\section{Discussion}
\par The Kamiokande
collaboration\cite{bib:Hirata} has reported a limit for 
$\pnk$ of $10.0 \times 10^{31} $ years
with 9 candidate events and an estimated background of 7.3 events.
Their unsubtracted limit is $5.1 
\times 10^{31}$ years.  IMB has reported a limit without 
background subtraction of 
$1.0 \times 10^{31}$ years  with 6 candidate events and an 
estimated
background of 4.7.  More recently,
they report $5.0 \times 10^{31}$ years  without a background subtraction based on
14 candidates\cite{bib:mcgrew}.
With background 
subtraction of 21.4 events, they calculated their limit to be $15.1 \times 10^{31} 
$ years.
Frejus has reported a background subtracted 
limit of $1.5 \times 10^{31}$ years  with 1 candidate and an 
estimated background of 
1.8.  In all experiments, the largest 
contribution to the estimated 
background comes from $\nu_\mu$ quasielastic scattering.  The 
existence of the atmospheric neutrino $\nu_\mu$ deficit then calls into
question the reliability of background estimates in all of these 
experiments.  It is not 
clear that rescaling the total $\nu_\mu$ rate to the observed rate 
leads to a correct estimate of the background for proton decay in 
this channel, since whatever is causing the atmospheric $\nu_\mu$ 
deficit may be energy dependent or different for $\nu_\mu$ and 
$\bar{\nu}_\mu$.  We note the possibility of such differences by
comparing plots 2b and 2c.  
We would assign a resulting systematic error of 
as much as 100\% to the 
background 
estimates for this channel.  In view of this large
systematic error we distrust the background subtraction in this channel
and prefer to rely on the unsubtracted limit, which is unaffected by the
background uncertainly.  We note that the background level  in this analysis, 
and thus the importance of background subtraction,
is an order of magnitude lower than in the water cherenkov experiments.
\par In summary, we have searched for the proton decay mode $\pnk$
using two decay modes of the $K^+$, $\km$ and $\kpz$.  We observe
one candidate event for $\km$ and  zero candidates for $\kpz$; 
the estimated backgrounds are 0.40 events and 1.14 events 
respectively.  
Our combined lower lifetime limit at 90\% CL is 
$4.3 \times 10^{31}$ years.   Our limit with background 
subtraction is $4.6 \times 10^{31}$ years.

\begin{ack}  This work was undertaken with the support of 
the U.S. Department of Energy, the State and University of
Minnesota and the U.K. Particle Physics and
Astronomy Research Council.
 We wish to thank the following for their invaluable help with      
the Soudan 2 experiment: the staffs of the collaborating laboratories; the
Minnesota Department of Natural Resources for allowing us to use the facilities
of the Soudan Underground Mine State Park; the staff of the Park, particularly
Park Managers D. Logan and P. Wannarka, for their day to day support; and Messrs
B. Anderson, J. Beaty, G. Benson, D. Carlson, J. Eininger and J. Meier of the
Soudan Mine Crew for their work in the installation and running of the
experiment.
\end{ack}


\begin{thebibliography}{99}
\bibitem{bib:sak1} A.D. Sakharov, Zh. Eksp. Teor. Fiz. Pis'ma {
5} (1967) 32 (JETP Lett. {5} (1967) 24).
\bibitem{bib:Hirata} K.S. Hirata \etal, {\it Kamiokande 
collaboration}, Phys. Lett. {B220} 
(1989) 308.
\bibitem{bib:becker} R. Becker-Szendy \etal, {\it IMB 
collaboration}, Phys. Rev. {D42} 
(1990) 2974.
\bibitem{bib:hisano} J. Hisano, H. Murayama, and T. Yanagida, Nucl. 
Phys. {B402} (1993) 46.
\bibitem{bib:wess} J. Wess and B. Zumino, Nucl. Phys. {B70} 
(1974) 39; Phys. Lett. {B49} (1974) 52.
\bibitem{bib:nath} P. Nath, A.H. Chamseddine, and R. Arnowitt, 
Phys. Rev. {D32} (1985) 2348.
\bibitem{bib:Haines} T. Haines \etal, {\it IMB collaboration},
Phys. Rev. Lett. {57}
(1986) 1986. 
\bibitem{bib:Berger} Ch. Berger \etal, {\it Frejus collaboration},
Nucl. Phys. {B313} 
(1989) 509.
\bibitem{modcon} W.W.M. Allison \etal,  Nucl. Instr. Meth. {A376} (1996) 36.
\bibitem{modperf} W.W.M. Allison \etal, Nucl. Instr. Meth. 
{A381} (1996) 385.
\bibitem{gallagher} H. Gallagher,  Neutrino Oscillation Searches with the
Soudan~2 Detector,  PhD Thesis, University of Minnesota 1996.
\bibitem{shield} W.P. Oliver \etal, Nucl. Instr. Meth. A 276 (1989) 371.
\bibitem{bib:allison} W.W.M. Allison \etal, Phys. Lett.  
{B391}, (1997) 491.
\bibitem{barr} G. Barr, T.K. Gaisser and T. Stanev, Phys. Rev. 
{D39} (1989) 3532. 
\bibitem{bodek} A. Bodek and J. Ritchie, Phys Rev. {D23} (1981) 
1070.
\bibitem{intranuke} W.A. Mann \etal, INTRANUKE: A phenomenological 
code for pion rescattering within an extended nucleus, Soudan 2 
internal note, PDK-377 (1988), unpublished; W. Leeson \etal,
Nucleon decay final state effects from Fermi motion and from 
INTRANUKE, Soudan 2 internal note PDK-678 (1997), unpublished.
\bibitem{merenyi} R. Merenyi \etal, Phys. Rev. {D45} (1992) 743 and
R. Merenyi, A Study of Intranuclear Scattering in $\nu_\mu$ Ne versus 
$\nu_\mu$ D Interactions: Implications for Nucleon Decay Searches, PhD Thesis,
Tufts University, 1990. 
\bibitem{garcia} C. Garcia-Garcia, El Experimento de Soudan 
2 para el Estudio de la Establilidad de la Materia:  Interacciones 
de Neutrinos, Ph.D. thesis, Universidad de 
Valencia, 1990.
\bibitem{capri} H.R. Gallagher, talk to appear in proceedings of 
WIN'97, Capri, Italy, June 1997; T. Kafka, talk to appear in 
proceedings of TAUP 97, LNGS, Assergi, Italy, September 1997.
\bibitem{bib:mcgrew} C. McGrew \etal, {\it IMB collaboration}, 
submitted to Physical 
Review, 1997.
\end{thebibliography}
\end{document}